\pdfoutput=1

\documentclass[11pt]{article}

\usepackage{acl}

\usepackage{times}
\usepackage{latexsym}
\usepackage{amsmath,amssymb}  
\usepackage[mathscr]{eucal}
\usepackage{graphicx}
\usepackage{enumitem}

\usepackage[T1]{fontenc}

\usepackage[utf8]{inputenc}

\usepackage{microtype}

\title{``Does it come in black?''\\CLIP-like models are zero-shot recommenders}

\author{Patrick John Chia\thanks{* GradRECS started as a (failed) experiment by JT; PC actually made it work, and he is the lead researcher on the project. FB, CG and DC all contributed to the paper, providing support for modelling, industry context and domain knowledge. PC and JT are the corresponding authors.} \\
  Coveo, Montreal \\
  \texttt{pchia@coveo.com} \\\And
  Jacopo Tagliabue \\
  Coveo Labs, New York \\
  \texttt{jtagliabue@coveo.com} \\\AND
  Federico Bianchi \\
  Bocconi University, Milan \\ \\\And
  Ciro Greco \\
  Coveo Labs, New York \\\\\And
  Diogo Goncalves \\
  Farfetch, Porto \\
  }

\begin{document}
\maketitle
\begin{abstract}
Product discovery is a crucial component for online shopping. However, item-to-item recommendations today do not allow users to explore changes along selected dimensions: given a query item, can a model suggest something similar \textit{but} in a different color? We consider item recommendations of the \textit{comparative} nature (e.g. ``something darker'') and show how CLIP-based models can support this use case in a zero-shot manner. Leveraging a large model built for fashion, we introduce \texttt{GradREC} and its industry potential, and offer a first rounded assessment of its strength and weaknesses.
\end{abstract}

\section{Introduction}

\begin{figure}
  \centering
  \hspace{-2.5mm}
  \includegraphics[scale=0.23]{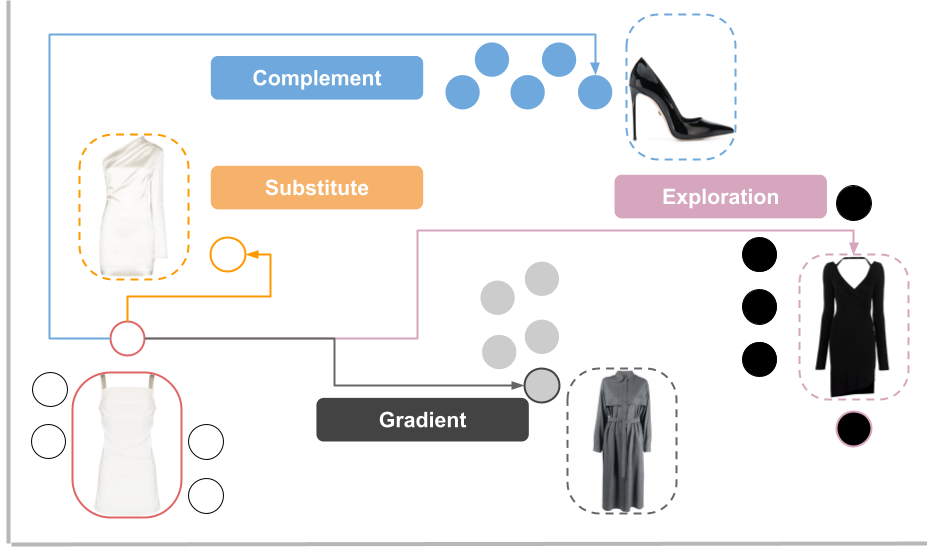}
  \caption{Recommendation as movements in the latent space, starting from a query item (\textit{red}): substitute, complementary and exploration-based strategies are depicted in orange, blue and violet; \texttt{GradREC} is in gray.}
  \label{fig:space}
  \vspace{-4mm}
\end{figure}

Recommender systems (RSs) are one of the most ubiquitous applications of machine learning (ML) in e-commerce \cite{Tsagkias2020ChallengesAR}, recently featuring novel benchmarks and extensive use of deep neural networks in item-to-item, user-to-item, and comparison RSs~\cite{Moreira2019OnTI,areUSure,CoveoSIGIR2021}. While details differ between neural architectures, they all share the principle that products are represented as points in a latent space, learned from user behavior, item meta-data or a combination of both~\cite{tagliabue-etal-2021-bert,48840}. Fig.~\ref{fig:space} represents item-to-item recommendations \textit{as movements in the product space}: starting from a query item -- the \textit{white dress} --, RSs help shoppers to move either around their current location, or ``jump'' to a different one. Adding to the blooming literature on substitute, complementary, popularity and exploration-based  RSs~\cite{10.1145/3397271.3401042,10.1145/3340531.3412732,10.1145/3383313.3411555,10.1145/3109859.3109866}, \textit{this} work presents \texttt{GradREC}, a new type of recommendation that introduces explicit directionality into the mix, by allowing exploration in selected directions \textit{through natural language}: ``something darker'' will move the user from the \textit{white dress} to the \textit{grey dress}. In particular, we summarize our contributions as follows: \textbf{First}, we introduce \texttt{GradREC} as a new type of recommendation experience \textit{and} a technical contribution -- to the best of our knowledge, \texttt{GradREC} is the first \textit{zero-shot} approach for language-based comparative recommendations, showing that CLIP-like \cite{Radford2021LearningTV} models may enable recommendations to be generated on the fly \textit{without the need of explicitly defined labels for training} or behavioral data. \textbf{Second}, we devise both qualitative and quantitative evaluations to offer a first rounded assessment of the strengths and weaknesses of our proposal, and supplement our analysis with extensive visual examples. \textbf{Third}, as part of our submission, we release to the community our fine-tuned weights, publish an interactive web-app for exploration, and open source our code to help reproducing our findings and building on them \footnote{Artifacts are available at \url{https://github.com/patrickjohncyh/gradient-recs}.}.

While we present our results as a \textit{preliminary} investigation into the untapped capabilities of CLIP for retail, we \textit{do} believe our methods to be interesting to a broad set of practitioners: those exploring recommendations for conversational and interactive commerce, and those leveraging deep learning for horizontally scalable SaaS products\footnote{As a context for this global market, Algolia and Bloomreach both raised more than USD200M in the last two years ~\cite{AlgoliaRound,BloomreachRound}, and Coveo raised more than CAD200M with its IPO \cite{CoveoRound}.}. Finally, while motivated by very practical concerns, this work contains new insights on the topology of the information encoded by over-parameterized neural networks, which could help our understanding of the kind of regularities that these models learn about our world.

\section{An Industry Perspective}

The intersection of product recommendation and natural language is a blooming research area: advances in neural NLP have been recently used for content-based recommendations~\cite{DBLP:conf/sigir/IqbalKA18}, cold-start scenarios~\cite{10.1145/3383313.3411477}, language grounding~\cite{bianchi-etal-2021-query2prod2vec,bianchi-etal-2021-language}, and explainable RSs \cite{DBLP:journals/corr/abs-2101-03392}. A relatively new use case is provided by the growth in the market of interactive technologies, as intelligent virtual assistants (IVAs) are expected to handle recommendations that increasingly encompass the expressiveness of natural language \cite{10.1145/3453154}. While interaction is an opportunity, the limited real estate available to display recommendations is a constraint for IVAs \cite{lin-etal-2021-personalized}: since scrolling is limited, strategies for moving from one product to another (as in Fig.~\ref{fig:space}) are crucial for IVAs market penetration. In \textit{this} work, we consider recommendations which are of the \textit{comparative} form: given an item of focus -- in a chat, a product page, etc. --, the shopper makes use of natural language queries to retrieve a second item (e.g. ``shorter``, ``darker'', etc.), related to the first but different along the specified attribute. While state-of-the-art IVAs can already provide very simple recommendations through language~\cite{StyleByAlexa}, we are the first to suggest the existence of an entire new dimension and depth to mimic the interactions typical of a real-life shopping experience. 

When thinking about applying this method in a multi-tenant SaaS context, it is worth noting how small are the assumptions \texttt{GradREC} actually makes about the underlying inventory: while in the case of \texttt{FashionCLIP} and its dataset it is true that products often contain information about an attribute's intensity (e.g. ``knee-length shorts''), the \textit{relationship} between them is not explicitly encoded, yet it is inferred by \texttt{GradREC}. Moreover, when applying these models across new catalogs, there is no guarantee descriptions would be as rich, or even using the same lexicon to describe the same attribute (``bermudas'' vs ``knee-length''). These considerations further highlight the strength of using a latent space derived from a general and flexible multi-modal model, and the non-trivial nature of extracting comparative recommendations. 

\section{Related Work}
Our work sits at the intersection of various recent technical advances in \textit{latent space manipulation} and \textit{iterative IR}. Many recent works explore latent space manipulation of Generative Adversarial Networks (GANs) for purposes of fine-grained image editing \cite{shen2020interpreting, Patashnik_2021_ICCV}; \citet{gansteerability} also studied latent space traversal in GANs to measure GAN generalization. We extend this line of research by providing a clear e-commerce use case, a focus shift from generative modeling to recommendation, and new insights on CLIP-based representations. 

The idea of iterative search refinement using comparative information and attribute ranking is not new \cite{whittlesearch,7410635}. However, previous work sit in the standard \textit{fully supervised} ``learning-to-rank'' tradition. Conversely, our approach operates in a zero-shot fashion by using \textit{both} CLIP retrieval and CLIP representations to generate suggestions on-the-fly. Finally, our work builds on top of the recent wave of contrastive-based methods for representational learning: while latent product representations have been extensively studied from multiple angles \cite{Bianchi2020FantasticEA,10.1145/3336191.3371778}, CLIP-like models are still very new in this domain: \textit{GradREC} leverages the space learned by \texttt{FashionCLIP}, a fashion-fine tuning of the original CLIP~\cite{underR}.

\section{Gradient Recs}

\subsection{Overview}
\label{sec:overview}

\texttt{GradREC}, builds upon the multi-modal space induced by \texttt{FashionCLIP}. \texttt{GradREC} aims to traverse the latent space such that the intensity of an attribute of interest varies monotonically for products along that path, allowing us to make fine-grained recommendations that require comparative knowledge. There are independently grounded reasons to expect this method to work. \textit{First}, we have solid evidence that embedding spaces are able to encode recognizable ``concepts'' (e.g. lexical knowledge in \textit{word2vec} \cite{mikolov-etal-2013-linguistic}, facial expressions in GANs \cite{ding2017exprgan}). \textit{Second}, we perform an extensive evaluation of the \texttt{FashionCLIP} product space, focusing on attributes such as \textit{color} and \textit{occasion}: our qualitative assessment (Section~\ref{appendix:tsne}) verified that embeddings are indeed often clustered, further suggesting that movements in the ``concept space'' can be represented as paths in the latent space.

\subsection{FashionCLIP exploration}
\label{appendix:tsne}
\begin{figure}[h]
  \centering
  \includegraphics[scale=0.115]{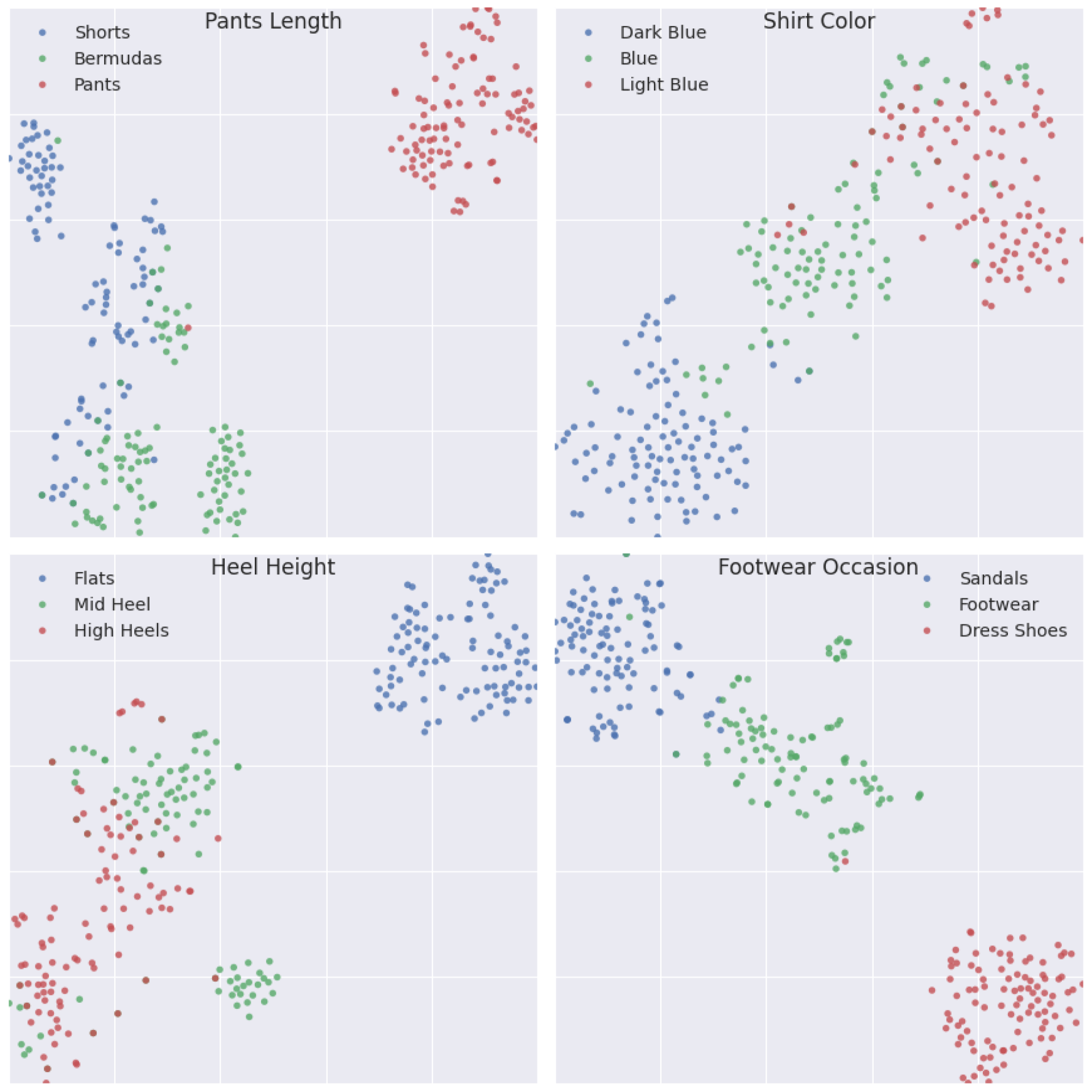}
  \caption{Sample TSNE projections of product image vectors: products are colored based on attribute strength.}
  \label{fig:tsne}
  \vspace{-4mm}
\end{figure}

As discussed in Section~\ref{sec:overview}, there are pre-existing theoretical reasons to think that embedding spaces encode in their geometry interesting regularities. In order to validate this hypothesis, we run visual investigations on \texttt{FashionCLIP} space as seen in Figure \ref{fig:tsne}, which shows TSNE projections of product image embeddings for four attributes: Pants Length, Shirt Color, Heel Height and Occasion. For each attribute, we retrieve products possessing negative, neutral and positive attribute intensities. Figure \ref{fig:tsne} demonstrates that the projected products from the corresponding attribute intensities do indeed form meaningful clusters, suggesting that it is possible to trace a path from one cluster to another in the latent space.

\subsection{Method}
\label{sec:method}

In what follows, we focus on the core task of gradient recommendations\footnote{We realize that a more ecological setting -- such as IVA -- would require additional steps to handle stateful interactions: those steps are however general open problems in IVA, whose solution is independent of the interaction we model here.}. Assuming a target inventory of fashion products, a starting item and a pair of natural language queries whose difference captures the comparative dimension of interest (e.g. the difference between ``dark red shirt'' and ``red shirt'' captures the dimension of ``darker''\footnote{Different ecological settings may provide these queries more or less explicitly; \texttt{GradREC} may be used naturally in the context of multi-turn systems such as IVAs, or, for example, as support to standard manually defined facets for IR use cases, such as product search.}), \texttt{GradREC} should return a new item in the ``same style'' as the starting item, varying along the specified dimension; in particular, \texttt{GradREC} can leverage  CLIP representations but has no access to labels or co-purchasing data. Providing now a formal description, we decompose our approach into two components: a \textit{traversal function}, $\mathbf{\Phi}$ and a \textit{traversal direction vector}, $\mathbf{v}_c$.

\textbf{Traversal Function}: given a product $t$, represented in the CLIP space by either its L2 normalized textual vector $\mathbf{t}_t$ or image vector $\mathbf{i}_t$, and some attribute $c$ we want to explore, our goal is to compute a function $\mathbf{\boldsymbol{\Phi}}$, such that given a starting point $\mathbf{v}_t$ and some vector $\mathbf{v}_c$, returns a new point $\mathbf{v}_{t+1}$ in the latent space that is increasing or decreasing in strength of attribute $c$. Given the new position $\mathbf{v}_{t+1}$, we use cosine-based k-nearest neighbors ($KNN(\cdot,\cdot)$) to retrieve suggested products: if we iterate this process, we would travel along the dimension of attribute $c$, discovering products as we move along. We define $\boldsymbol{\Phi}$ as vector addition, with a scale factor $\lambda$ to control step size; additionally, we use the mean of the current point's nearest neighbours ($\bar{KNN}(\mathbf{v}_t, k)$) as a regularizing term. The two terms are balanced by taking a convex combination of the direction vector and the regularizing term. In our notation, $\mathbf{\hat{v}}$ refers to $\mathbf{v}$ normalized to unit length. Note that all vectors are of dimension 512. Our definition is summarized in Eq.\ref{eq:main_eqn}:
\vspace{-5mm}

\begin{equation}
\label{eq:main_eqn}
    \begin{split}
    \mathbf{v}_{t+1} = & \;\boldsymbol{\Phi}(\mathbf{v}_t, \mathbf{v}_c)\\
                     = & \;\mathbf{v}_t + \;(1-\rho)\cdot\lambda\mathbf{\hat{v}}_c\\
                       & \quad\; +\rho\cdot{K\bar{N}}N(\mathbf{v}_t, k)
    \end{split}
    \vspace{-8mm}
\end{equation}

\textbf{Traversal Vector}: the construction of $\mathbf{v}_c$ relies on two main ingredients. First, given a pair of queries which semantically captures the attribute $c$ (``darker''), we use the zero-shot retrieval capabilities of \texttt{FashionCLIP} to construct two small datasets: one comprising the image embeddings closest to the \texttt{FashionCLIP} encoding of the neutral class (``a blue shirt''), and one from an exemplar class for $c$ (``a dark blue shirt''). We define the retrieved image vectors for the neutral and exemplar prompts as $\mathbf{I}_n = \{\mathbf{i}_n^1 ... \mathbf{i}_n^M\}$ and $\mathbf{I}_e = \{\mathbf{i}_e^1 ... \mathbf{i}_e^N\}$ respectively. Second, we adopt the channel importance measure from \citet{DBLP:conf/cvpr/WuLS21} to determine channels\footnote{Each channel corresponds to one of the 512 dimensions of an embedding.} which encode the differences between the neutral class and exemplar class. The method measures the channel-wise Signal-to-Noise Ratio (SNR) between the mean neutral class vector (i.e. $\mathbf{\bar{I}}_n$) and the exemplar class vectors (i.e. $\{\mathbf{i}_e^1 ... \mathbf{i}_e^N\}$). The intuition is that channels with high SNR correspond to channels which encode the differences between images from the neutral and exemplar class, and hence the attribute $c$. Our implementation departs from theirs by retaining the sign of the differences for each channel. Finally, to obtain $\mathbf{v}_c$, we normalize the vector formed by the channel-wise SNR values\footnote{We refer the reader to \citet{DBLP:conf/cvpr/WuLS21} for the original discussion.}.

\begin{figure}
  \centering
  \includegraphics[scale=0.25]{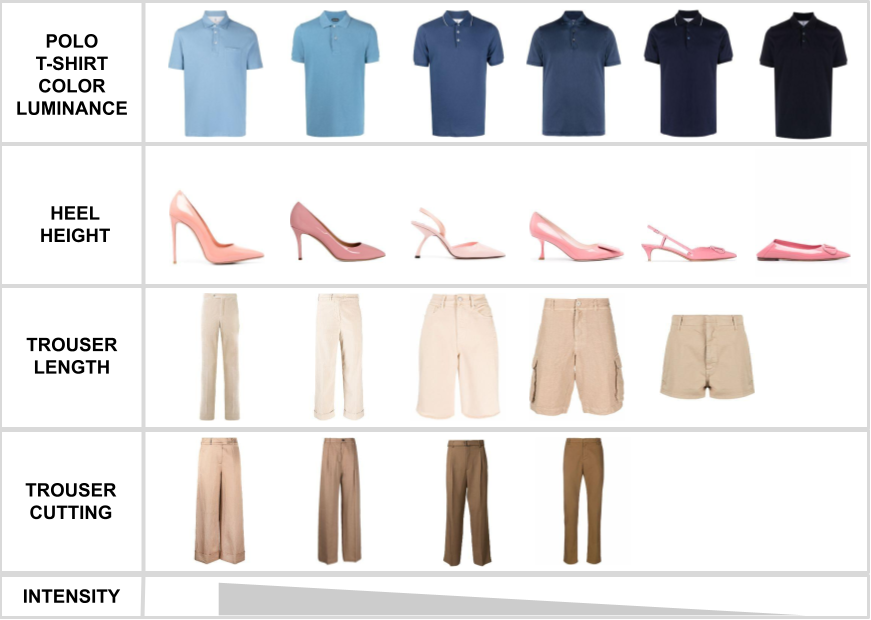}
  \caption{A sample of the qualitative results obtained by applying \texttt{GradREC} for four different attributes: the \textit{intensity} / \textit{strength} of the attribute decreases from left to right.} 
  \label{fig:examples}
  \vspace{-4mm}
\end{figure}

\section{Experiments}

To investigate \texttt{GradREC} strengths and weakness, we offer a preliminary assessment of its capabilities over important fashion dimensions, such as product discovery.

\subsection{Dataset and Pre-trained Space}
Our pre-trained space is \texttt{FashionCLIP}, an adaptation of CLIP obtained by fine-tuning the original embeddings over fashion products provided by \textit{Farfetch}, a world leading platform for online luxury fashion shopping. The dataset comprises of over 800k fashion products across dozens of item types and more than 3k brands. In addition to a standard product image over white background, the dataset contains natural language descriptions of the stylistic properties (e.g., ``cotton-blend'', ``high waist'',  ``belt loops'') and categorical information (e.g. ``layered track shorts'') of products\footnote{\texttt{FashionCLIP} weights and training code will be released with the original publication. At the moment of writing this paper, the original training dataset is scheduled to be released as well: please check \url{https://github.com/Farfetch} for updates.}. \texttt{FashionCLIP} shares the same architecture as~\citet{Radford2021LearningTV}, i.e. a multi-modal model comprising an image and a text encoder. We refer to \citet{underR} for details on training and retrieval / classification capabilities: since \texttt{FashionCLIP} has independent value in the industry, \texttt{GradREC} does not require any \textit{specific} pre-training.

\subsection{Qualitative Analysis}

We consider four different attributes of interest: \textit{shirt color luminance}, \textit{heel height}, \textit{trouser length} and \textit{trouser cutting}. For each attribute, we traverse the latent space between both extremes of the attribute of interest and present the results in Fig.~\ref{fig:examples} for visual validation. We observe that the products retrieved form a monotonic change in the attribute's strength that aligns well with human intuition: i.e. t-shirts in the first row do indeed follow a gradient going from lighter to darker shades of blue. It is interesting to note that the latent space of \texttt{FashionCLIP} appears to encode and organize these geometric and physical regularities despite not having been trained to do so explicitly, pointing to further questions about what and how these self-supervised models learn. 

\subsection{Quantitative Analysis}

\begin{figure}[h!]
  \includegraphics[clip,trim=0 255 130 0, scale=0.30,width=\columnwidth]{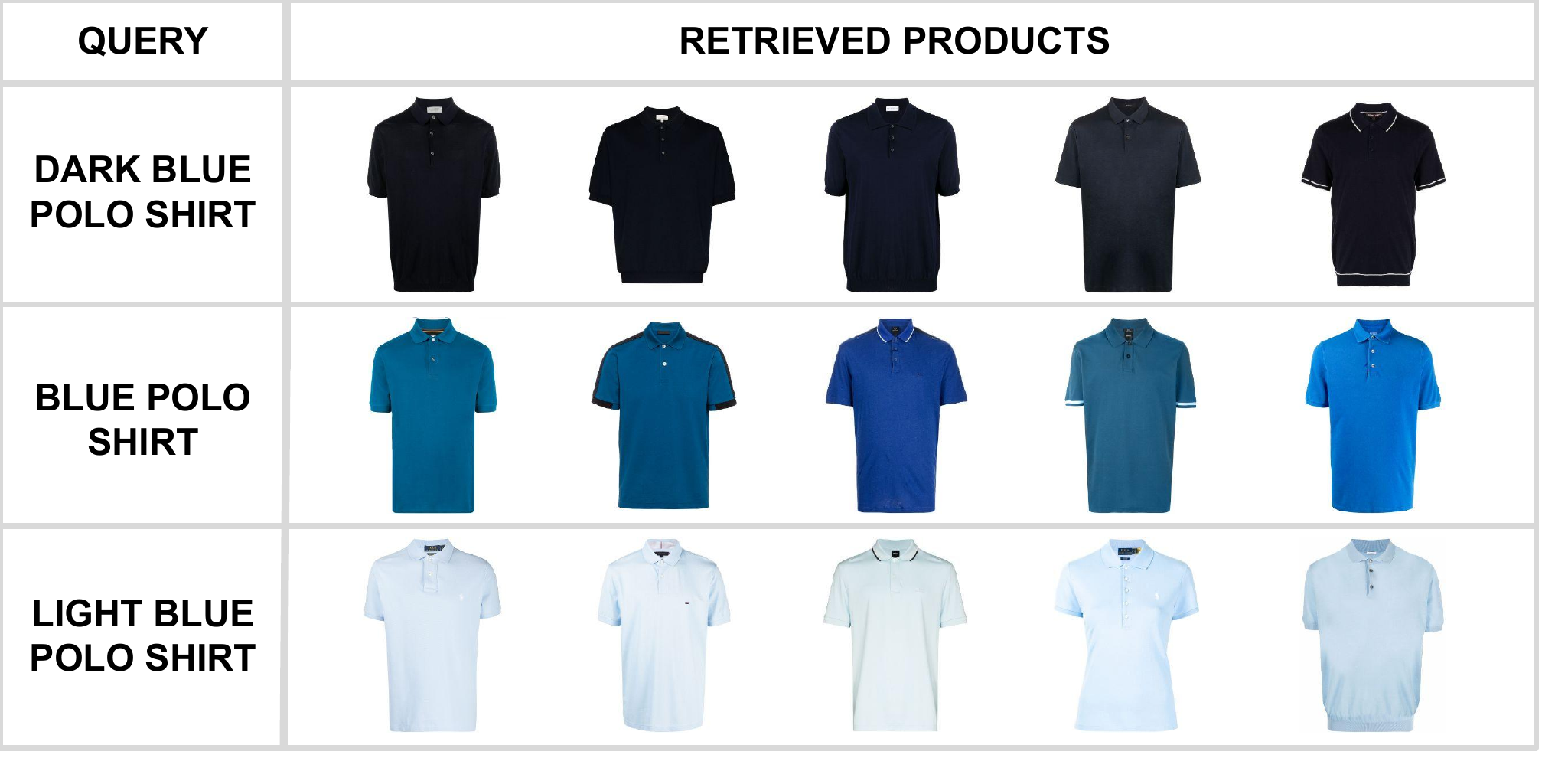}
  \caption{Products retrieved for queries on a spectrum of intensity.}
  \label{fig:query_sample}
  \vspace{-4mm}
\end{figure}

We quantitatively assess \texttt{GradREC} by measuring its efficiency in product discovery along an attribute of interest, to verify that the path it discovers is semantically meaningful. We generate three datasets ($N = 100$) using \texttt{FashionCLIP} retrieval capabilities to represent products from the negative, neutral and positive intensities of an attribute\footnote{The exact definitions of negative and positive are relative; We are more concerned here with capturing the opposing extremes of an attribute's spectrum i.e. \textit{dark} and \textit{light}.}. For example, to generate datasets for shirt color \textit{luminance} we would issue the following queries -- ``dark blue polo shirt'', ``blue polo shirt'', ``light blue polo shirt'' -- to \texttt{FashionCLIP} and retrieve $N$ products for each query. In Figure~\ref{fig:query_sample} we visualize a sample of the products retrieved for each of the above queries.

We apply \texttt{GradREC}, starting a traversal from a negative product in the direction of neutral intensity products, simulating product discovery by logging the top $k=10$ unseen products found at each step. We then compute the intersection cardinality of the three datasets along the simulated trajectory with a sliding window of 50 products: a model which traverses a meaningful path should produce three peaks, one for each level of intensity. As a baseline, we use visual similarity in the CLIP image-space (KNN over image embeddings) and simulate the product discovery trajectory as traversing the list of nearest neighbors of the same seed product in the order of increasing distance.

\begin{figure}
  \includegraphics[scale=0.38]{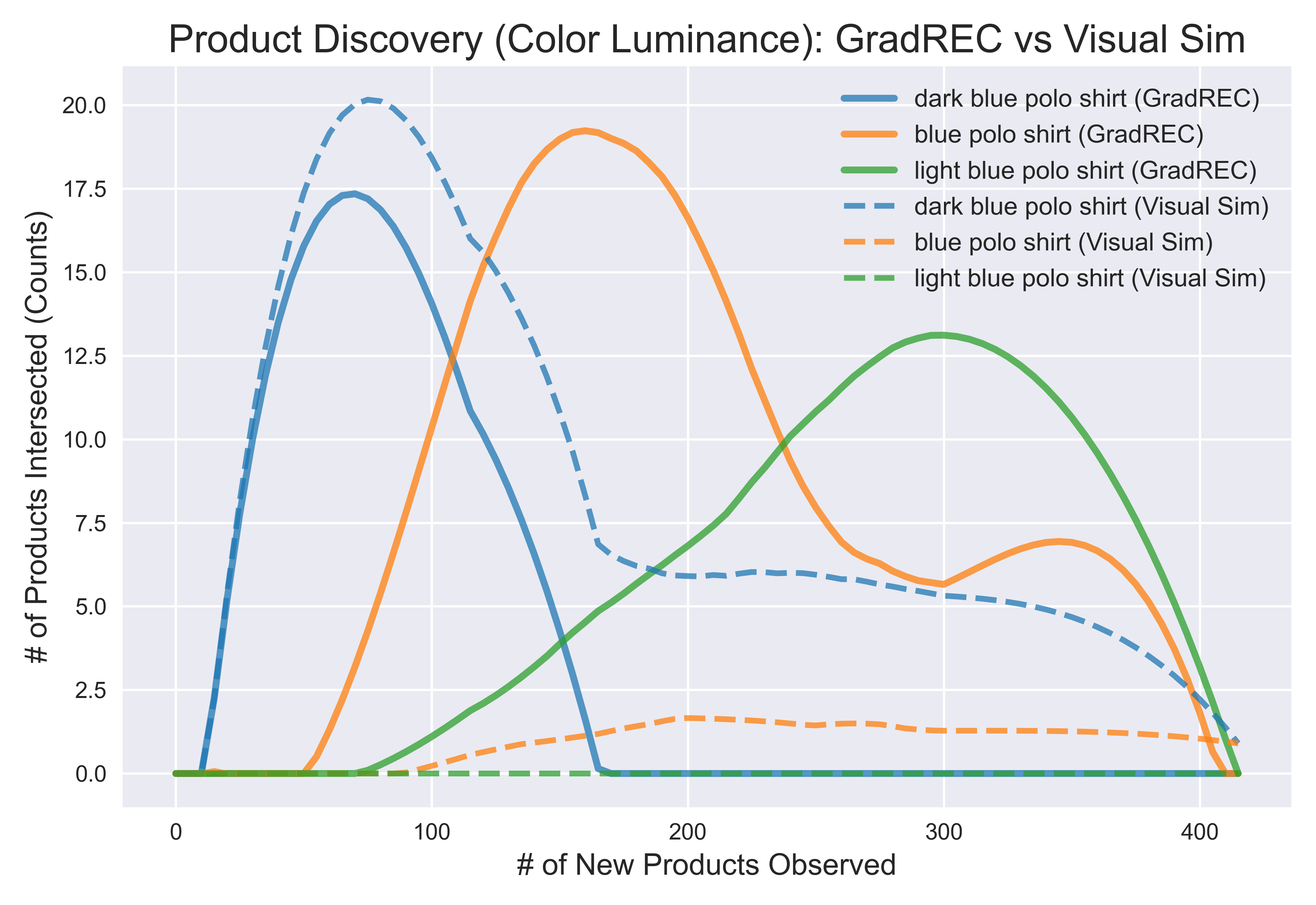}
  \caption{Quantitative analysis comparing \texttt{GradREC} to Visual Similiarty on product discovery for the attribute \textit{color luminance} of blue shirts.}
  \label{fig:eval_quant}
  \vspace{-5mm}
\end{figure}

In Figure \ref{fig:eval_quant} we see the result of applying our analysis to the discovery path for \textit{luminance} of blue polo shirts. We observe that \texttt{GradREC} explores well this path as seen by its three distinct modes of intersected products, where each peak for \textit{light blue}, \textit{blue} and \textit{dark blue} respectively, corresponds to the correct \textit{order} of decreasing luminance. Conversely, we see that visual similarity fails to produce a similar product discovery pattern as \texttt{GradREC}, which spans a wider range of the luminance spectrum. In fact, visual similarity struggles to discover products from \textit{blue} and \textit{light blue}, highlighting the merits of the directionality induced by \texttt{GradREC} (Appendix \ref{appendix:exp}).

\subsection{Limitations \& Future Work}
\label{sec:lims}

While \texttt{GradREC} performances are encouraging -- especially when considering that no attribute has been explicitly taught --, limitations highlight several areas of improvement. First, the performance of the model is sensitive to the quality of the retrieval phase. For example, to construct $\mathbf{v}_c$ for \textit{trouser length}, using queries ``shorts'' and ``pants'' yielded better performance than ``shorts'' and ``bermudas''. Second, our definition of $\mathbf{\Phi}$ is not optimal: as we traverse the image space, the cosine distance between our position and all the products increases, suggesting that we are not traversing the latent manifold in the most efficient way. Third, while we observe that \texttt{GradREC} moves in a semantically meaningful direction, it does not, nor is it currently designed to, provide guarantees on the monotonicity of the products it returns along the path. Finally, \texttt{GradREC} does not account for uncertainty and thus does not possess a confidence measure for its recommendations: while we may be confident in its ability with geometric and physical concepts, and less so for more abstract notions (e.g. ``for colder weather''), it is hard to know \textit{a priori} what \texttt{GradREC} does not know.

\section{Conclusion}
We introduced \texttt{GradREC}, a zero-shot approach for comparative recommendations, that showed promising results in our initial investigations. While further evaluation -- especially, involving relevance judgments by humans -- is needed to fully assess \texttt{GradREC} capabilities, we \textit{do} believe that our work provides preliminary but novel insights into innovative application of large models in important industry use-cases.

\bibliography{anthology,custom}
\bibliographystyle{acl_natbib}

\appendix
\section{Appendix}



\subsection{A worked-out traversal example}
\label{appendix:traversal}

\begin{figure}[h]
  \hspace{-1mm}
  \includegraphics[scale=0.45, clip, trim=120 20 110 20,width=\columnwidth]{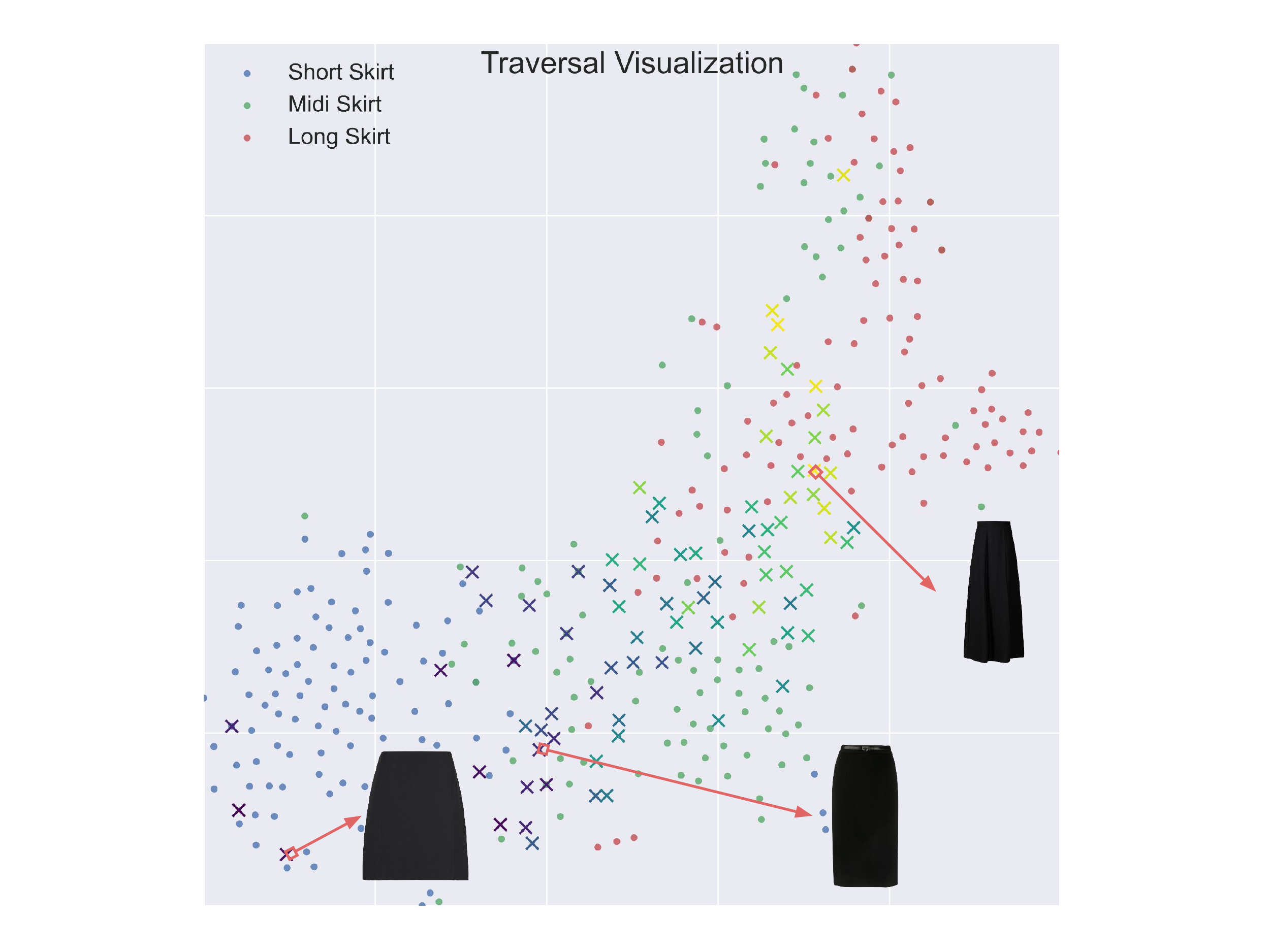}
  \caption{TSNE Projection of 3 ranges of skirt lengths and the traversed product path by \texttt{GradREC} along the attribute \textit{skirt length} as shown by $\times$ markers (dark to light denotes direction of traversal). Corresponding product images along the traversed path are visualized. }
  \label{fig:workedout}
  \vspace{-4mm}
\end{figure}

Fig.~\ref{fig:workedout} showcases an example of successful traversal when using the methods in Section~\ref{sec:method} applied to skirt length. In particular, we can see the query item, an intermediate one and a final one, along the path of products that was traced (the path of products are denoted by $\times$ markers, and the order of observation/direction of traversal is denoted by the darker to lighter hue). Note that a simple visual similarity search would have moved us from the first query item to a  nearby region.

\subsection{Additional Experiments}
\label{appendix:exp}

\begin{figure}[t]
  \includegraphics[scale=0.37]{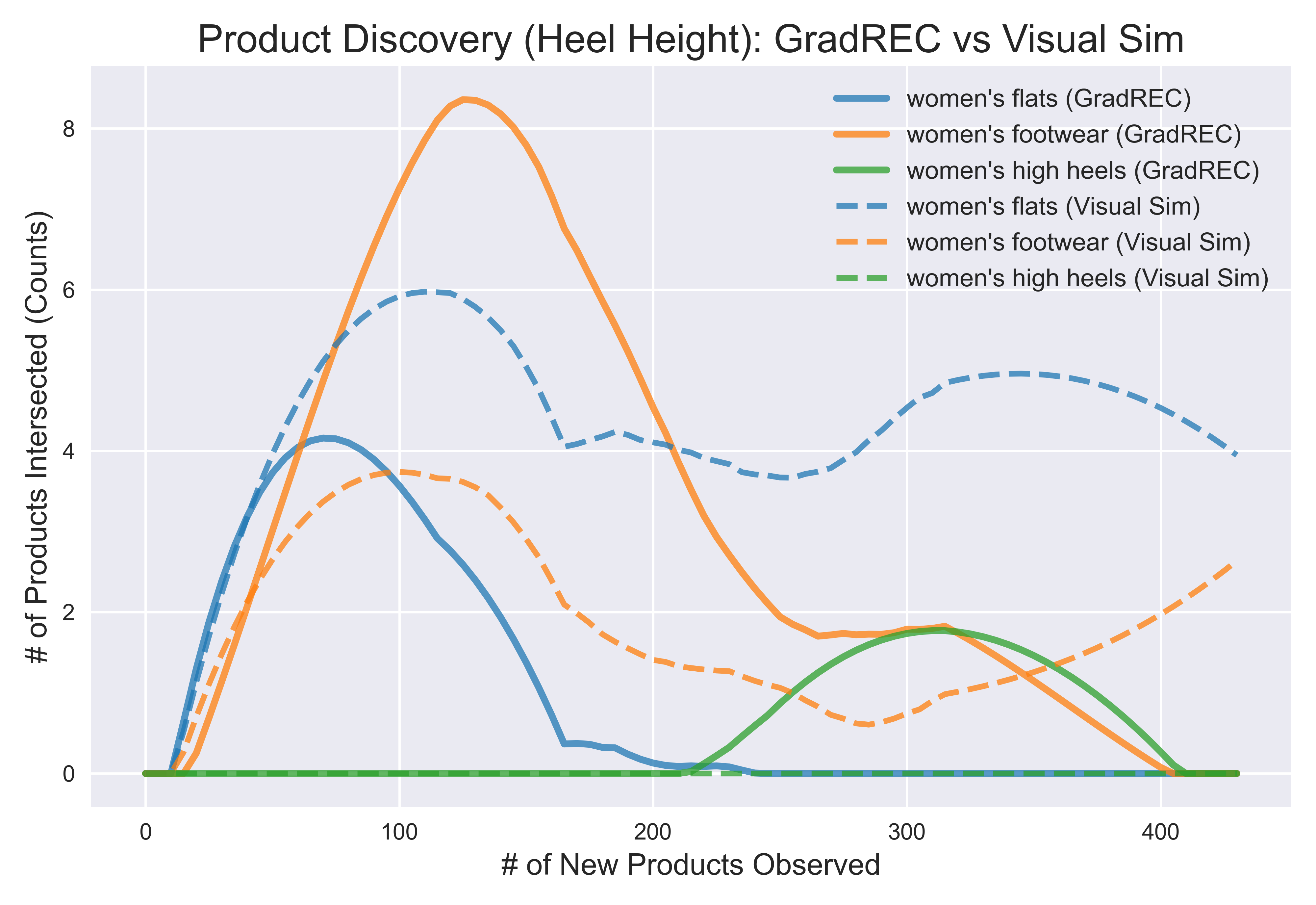}
  \caption{Product discovery analysis for \textit{heel height}.}
  \label{fig:quant_heel}
  \vspace{3mm}
\end{figure}

We ran our product discovery analysis for the attribute \textit{heel height} and report the result in Fig.~\ref{fig:quant_heel}.

A similar pattern as Fig.~\ref{fig:eval_quant} emerges with \texttt{GradREC} having three peaks and Visual Similarity struggling to discover ``high heel''. Unlike Fig.~\ref{fig:eval_quant}, however, we observe a lower cardinality of intersection for \texttt{GradREC} and ``women's high heels'', since \texttt{GradREC} preserves the style of the seed product (red colored shoes, in this example) while the products retrieved by ``women's high heels'' are of varying color.

\begin{figure}[h!]
  \hspace*{-2mm}  
  \includegraphics[clip,trim=0 170 0 0, scale=0.38]{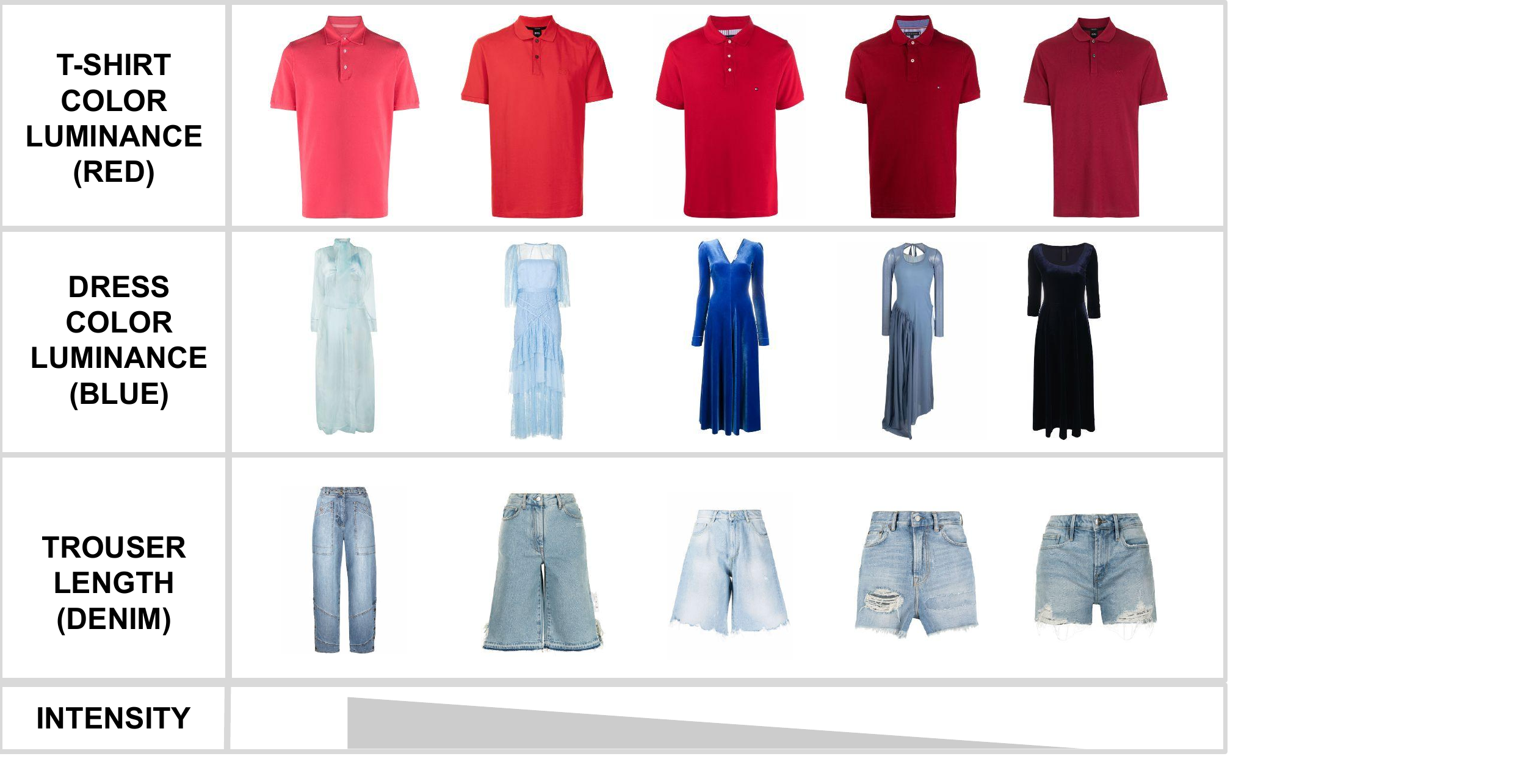}
  \caption{Extra qualitative examples.}
  \label{fig:egs_extra}
\end{figure}

We also provide additional qualitative examples in Fig.~\ref{fig:egs_extra}. We observe \texttt{GradREC} working across colors, in different product sortals (e.g., Dress), and having the ability to preserve visual style (e.g., Denim).

\begin{figure}[ht]
  \includegraphics[clip,trim=0 390 0 0,scale=0.35]{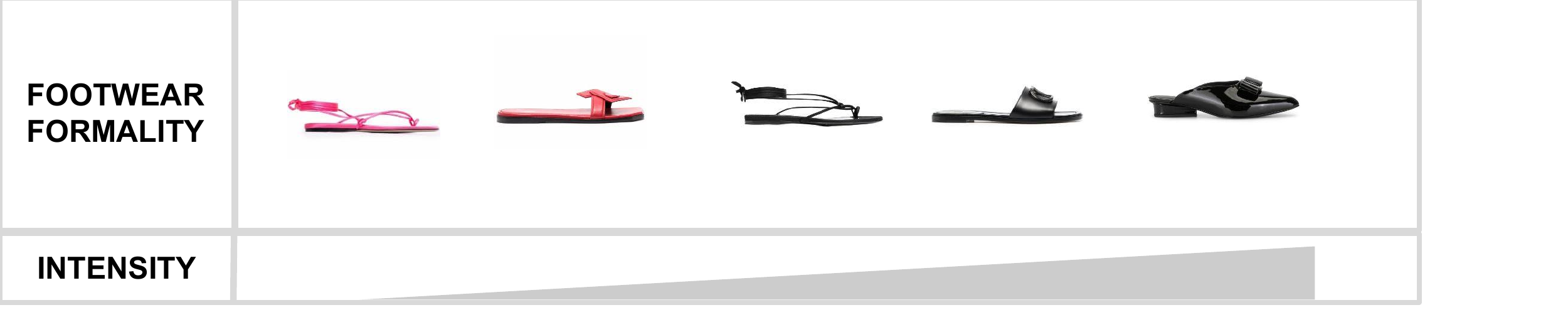}
  \caption{Failure mode of \texttt{GradREC} for \textit{formality} attribute. While \textit{formality} is appropriately increased, the product changes visual appearance from pink to black.}
  \label{fig:egs_fail}
\end{figure}

In Fig.~\ref{fig:egs_fail}, we instead give an example of the limitations of \texttt{GradREC}. Indeed, we see a failure mode where \texttt{GradREC}, while increasing the strength of \textit{formality} correctly, is however unable to preserve the visual style of the footwear correctly. As we have highlighted in Section~\ref{sec:lims}, \texttt{GradREC} performance is sensitive to the initial dataset retrieval performance: in this instance, the query ``formal shoes'' retrieves predominantly black, leather dress-shoes, thereby steering the traversal in that direction. 

\end{document}